\documentclass[preprint]{raa}  

\usepackage{graphicx,times}             
\usepackage{natbib}
\usepackage{amssymb,amsmath}
\usepackage[utf8]{inputenc}
\bibpunct{(}{)}{;}{a}{}{,}
\usepackage[pagebackref=true]{hyperref}

\newcommand{\orcid}[1]{%
    \raisebox{0.7ex}{\scalebox{1}{
        \href{https://orcid.org/#1}{\includegraphics[height=1.5ex]{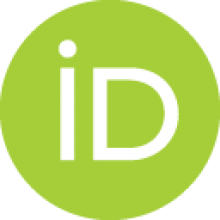}}%
    }}%
}

\newif\ifshowrevision
\showrevisiontrue  

\ifshowrevision
  
  \newcommand{\revdel}[1]{\textcolor{red}{\sout{#1}}}
  
\else
  \newcommand{\revdel}[1]{}     
\fi

\usepackage[labelfont=bf]{caption} 
\usepackage{xcolor}
\usepackage{tcolorbox} 
\tcbuselibrary{breakable} 
\newif\ifshowcomments
\showcommentstrue 

\ifshowcomments
  \newcommand{\highlightcomment}[2][yellow!20]{%
    \begin{tcolorbox}[
      colback=#1,            
      colframe=#1,           
      boxrule=0pt,           
      arc=0pt,               
      left=2pt,right=2pt,    
      top=1pt,bottom=1pt,    
      breakable              
    ]
    #2
    \end{tcolorbox}%
  }
\else
  \newcommand{\highlightcomment}[2][yellow!20]{} 
\fi

\usepackage{xcolor} 

\begin{document}

 \title{\textit{SVOM}/C-GFT: Instrumentation and Performances on the \textit{SVOM} Alerts}
  
   \volnopage{Vol.0 (202x) No.0, 000--000}      
   \setcounter{page}{1}          

\author{
Chao Wu\orcid{0009-0001-7024-3863}
\inst{1,2}
\and Zhe Kang
\inst{3,2}
\and Xiao-Meng Lu
\inst{1}
\and Xu-Hui Han\orcid{0000-0002-6107-0147}
\inst{1}
\and Li-Ping Xin\orcid{0000-0002-9422-3437}
\inst{1}
\and Pin-Pin Zhang
\inst{1}
\and You Lv
\inst{3}
\and Cheng-Wei Zhu
\inst{3}
\and Ruo-Song Zhang
\inst{1}
\and Jin-Song  Deng\orcid{0000-0001-5646-8583}
\inst{1,2}
\and Yu-Lei Qiu
\inst{1}
\and Mao-Hai Huang
\inst{1,2}
\and Hong-Bo Cai
\inst{1}
\and Hai-Bo Hu
\inst{1}
\and Lei Huang
\inst{1}
\and Lei  Jia
\inst{1}
\and Yu  Luo
\inst{1}
\and Jing Wang
\inst{1}
\and Mo Zhang
\inst{1}
\and Si-Cheng Zou
\inst{1}
\and Zhen-Wei Li
\inst{3}
\and Cheng-Zhi Liu
\inst{3}
\and Jian-Yan Wei
\inst{1,2}
}
\institute{
National Astronomical Observatories, Chinese Academy of Sciences, Beijing 100101,  China; 
{\it cwu@nao.cas.cn} {\it jsdeng@nao.cas.cn} \\
\and 
School of Astronomy and Space Science, University of Chinese Academy of Sciences, Beijing 101408, China \\ 
\and 
Changchun Observatory, National Astronomical Observatories, Chinese Academy of Sciences,Changchun 130117, China; \it{kangz@cho.ac.cn} \\ 
    {\small Received 202x month day; accepted 202x month day}}

\abstract{ 
The Chinese Ground Follow-up Telescope is an optical facility upgraded to support the Space Variable Objects Monitor (\textit{SVOM}) mission. Located at the Jilin Observation Station, it is capable of rapidly identifying and monitoring the optical counterparts of gamma-ray bursts. The 1.2-m telescope is equipped with two switchable focal-plane instruments: the prime-focus Large-FOV Transient Imager with CMOS and the Cassegrain-focus Camera with Three Channels. In this paper, we present a system overview, including the observatory, the telescope, the instrumentation, the automated operational framework managed by the Operation Center, and the data processing pipelines. We also report the performance results obtained over one year of \textit{SVOM}'s post-launch operations. The results demonstrate that the system meets its design specifications and delivers robust observational and operational performance. 
}

   \authorrunning{C. Wu, Z. Kang \& X.M. Lu }            
   \titlerunning{\textit{SVOM}/C-GFT INSTRUMENTATION AND PERFORMANCES}  

   \maketitle
             \section{Introduction}           
\label{sect:intro}

To unveil the nature of gamma-ray bursts (GRBs), one of the most luminous cosmic transients that are capable of releasing up to $\sim10^{49}-10^{55}\,\mathrm{erg}$ of isotropic equivalent energy in a brief prompt emission phase lasting from a few to tens of seconds \citep{Ghirlanda2004,Atteia2025}, the multi-wavelength follow-up of their relatively long-lived afterglows is crucial. It can constrain the energetics and environments of GRBs, and probe the physics of relativistic jets and shock interactions with the circumburst medium \citep{Zhangbingdoi:10.1142/S0217751X0401746X,Grenier2024A&A...691A.158G}. Specifically,  early-time optical observations, within the first minutes after the prompt emission, provide unique insights into the transition from prompt to afterglow phases, the respective roles of reverse and forward shocks, jet composition and magnetization, the microphysics of particle acceleration and radiation mechanisms, and the immediate environment of their progenitors \citep{kann2024A&A...686A..56K,Ramirez2024A&A...692A...3S,wang2024ApJ...969..146W}. Moreover, the rapid detection and fine localization of optical counterparts can promptly guide large optical/infrared telescopes for deeper observations, while early multi-band photometry enabling redshift estimation facilitates timely follow-up and spectroscopic identification of high-redshift GRBs.

A special system of Ground Follow-up Telescopes (GFTs) is designed for the Space Variable Objects Monitor (\textit{SVOM}) satellite, a joint Chinese-French GRB mission \citep{2016arXiv161006892W,ISCordier2026b}, to rapidly identify and monitor the optical counterparts of GRBs in the very early phases, complementing the onboard Visible Telescope (VT, \citealt{vt_overview_qiu+etal+2026}),
   which can begin observations only after the spacecraft completes the GRB-triggered slew and stabilizes on the target. 
Ground-based follow-up telescopes can often respond more rapidly, provided that the target is observable in the night sky, since their response time is shorter than the spacecraft slewing time. They can therefore fill the very early observational gap before VT observations begin. In addition, VT observations can also be triggered later through ground-commanded ToO observations, which typically have longer response times, further highlighting the importance of rapid ground-based follow-up.
This system consists of a Chinese GFT (C-GFT) and a French-Mexican GFT (FM-GFT, also known as COLIBR\'{I}; \citealt{2022SPIE12182E..1SB,svomissue-FGFT-Basa} ), which together form the core components of \textit{SVOM}’s global follow-up telescope network.
 
The C-GFT is a 1.2-m fast-response telescope at the Jilin Observation Station of the Changchun Observatory, NAOC, equipped with two instruments optimized for optical transient observations: the prime-focus  Large-FOV Transient Imager with CMOS (LATIOS) for wide-field monitoring, and the Cassegrain-focus Camera with Three Channels (CATCH) for simultaneous multi-band imaging. The telescope operates in synergy with the FM-GFT in Mexico, which also possesses additional near-infrared capabilities. Both telescopes respond promptly to the GRB localization alerts of the \textit{SVOM/ECLAIRs} \citep{ECL2025svom,ISSchanne2026} instrument, with some of these localizations further refined by the Microchannel X-ray Telescope \citep{ISGotz2026}, and together with the Ground-based Wide Angle Camera array (GWAC, \citealt{2021PASP..133f5001H, 2023NatAs...7..724X,svomissue-GWAC-xin}) and the sensitive onboard VT, provide an unprecedented temporal coverage of GRB optical counterparts, capable of capturing precursors, prompt flashes, and afterglows from the very early to late phases.

The structure of this paper is as follows. Section~\ref{sec:instrument-overview} provides an overview of the C-GFT system, including the observatory, the telescope, and the two focal-plane instruments, LATIOS and CATCH. It also describes the operational framework, covering the alert response and Operation Center, the telescope and observation control systems, and the data processing pipelines. 
Section~\ref{sec:performance-analysis} presents performance results from over one year of commissioning and subsequent routine operations. Finally, Section~\ref{sec:conclusion} offers a summary and outlook.

\section{The C-GFT SYSTEM}
\label{sec:instrument-overview}
\subsection{Telescope site}
\label{subsec:observatory}
 
The site of the C-GFT, the Jilin Observation Station of the Changchun Observatory, NAOC, is located at $126^{\circ}19'50''$,E, $43^{\circ}49'28''$,N, with an elevation of 320,m. It offers $\sim$230 observable nights per year and a median seeing of $\sim$1.3,arcsec \citep{KANG202412}. The observing environment exhibits strong seasonal variations: due to its relatively high latitude, the site benefits from long winter nights but experiences short summer nights. Winters are cold ($-20^{\circ}$C to $-30^{\circ}$C at night), dry, and generally clear, allowing for up to 13 hours of uninterrupted observations, whereas summers are humid with frequent rainfall, further limiting the already short observing window. Consequently, telescope maintenance is scheduled during summer. To support C-GFT operations, an unattended weather station and an all-sky cloud camera operate year-round, ensuring continuous environmental monitoring.

\subsection{Telescope}
\label{subsec:telescope}
 
\begin{figure}
  \centering
  \includegraphics[width=8cm, angle=0]{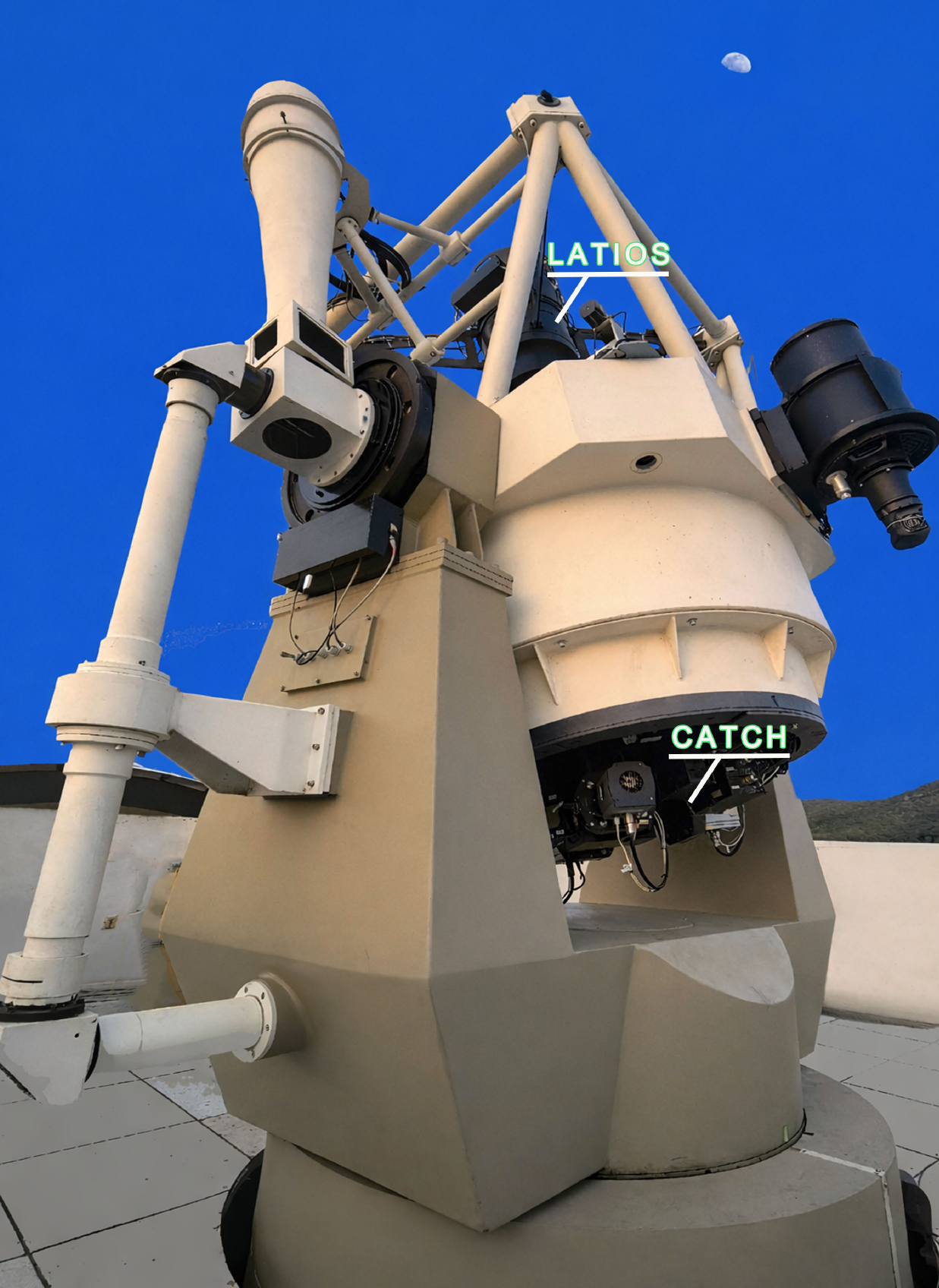}
  \caption{ A processed image of the C-GFT ready for observations. Two instruments mounted on the two focal systems (LATIOS and CATCH) are labeled directly in the figure for clarity.}
  \label{cgft_telescope}
  \end{figure}

\begin{figure}
  \centering
  \includegraphics[width=8cm, angle=0]{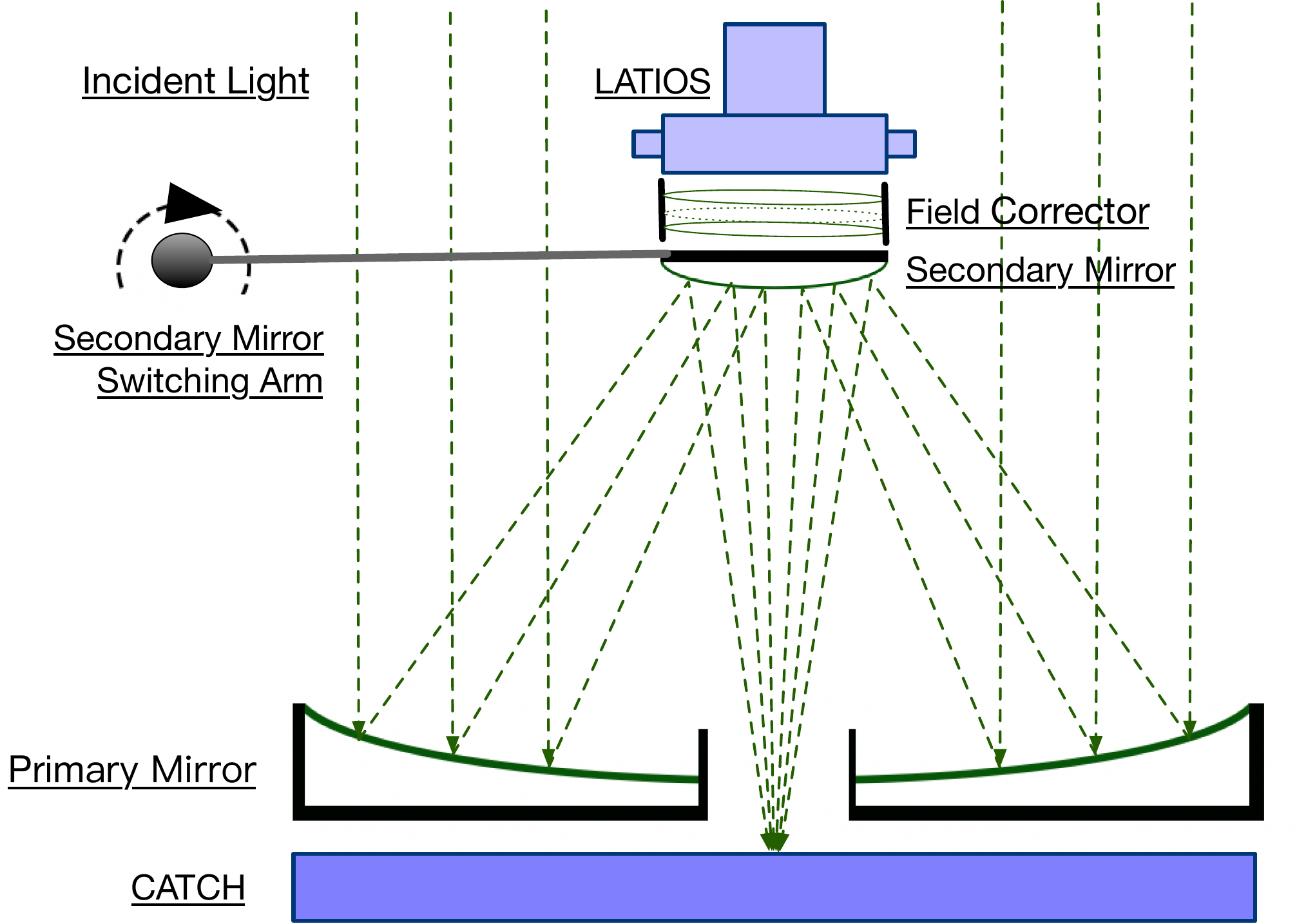}
  \caption{Optical layout of the C-GFT, showing the prime-focus LATIOS and Cassegrain-focus CATCH instruments, as well as the focus-switching mechanism.}
  \label{tel_struct_opticalpath}
  \end{figure}

The structure and optical layout of the telescope are shown in Figures~\ref{cgft_telescope} and~\ref{tel_struct_opticalpath}, respectively, with key parameters summarized in Table~\ref{tab:telescope_params}. It is a Ritchey–Chrétien (R–C) system with an aperture of 1.2 m and two mechanically switchable foci: the prime and Cassegrain.  The prime focus mounts the LATIOS instrument, which provides a wide $1.28^\circ \times 1.28^\circ$ field of view (FOV) at $f/1.83$ focal ratio, optimized for time-domain surveys of transient sources, particularly those with large localization uncertainties. The Cassegrain focus mounts the CATCH instrument, delivering an effective FOV of $21'\times21'$ at an $f/3.63$ focal ratio, with 80\% of the encircled energy (EE) contained within a $1.4''$ diameter. The focus switching is performed during daytime by manually engaging or retracting the secondary mirror.

The telescope features an alt-azimuth mount with high-precision friction drives, achieving a pointing accuracy better than $5''$ root mean square (RMS) per axis for zenith distances below $70^\circ$. The telescope's fast slewing capability, up to $15^\circ~\mathrm{s}^{-1}$ in azimuth, combined with a roll-off roof design which minimizes mechanical overhead by eliminating dome rotation delays, enables acquisition of new targets within 30--40~s.

Commissioned in 2017, the telescope has remained in stable operation for eight years. With the subsequent integration of the LATIOS and CATCH instruments under the \textit{SVOM} project, it now combines rapid response, wide-field capability, and high optical precision, making it an efficient facility for time-domain and photometric studies.

\begin{table*}[htbp]
\begin{center}
 \caption{Main Characteristics of the C-GFT.}
\label{tab:telescope_params}
\begin{tabular}{l l}
 \hline
\textbf{Parameter} & \textbf{Value} \\
\hline
 Effective aperture diameter & 1200~mm \\
    LATIOS field of view  & $1.28^{\circ} \times 1.28^{\circ}$ \\
LATIOS f-number  & 1.83   \\
Energy concentration at prime focus & 80\% enclosed within $0.75''$ \\
CATCH field of view  & $21' \times 21'$ \\
CATCH f-number  & 3.63   \\
 Energy concentration at CATCH & 80\% enclosed within $1.4''$ \\
     Zenith avoidance zone  & $2.5^{\circ}$ \\
Pointing accuracy (RA/Dec) & $5\arcsec$ rms \\
Tracking accuracy ($20^{\circ} < \mathrm{Alt} < 75^{\circ}$) & $0.2''$ rms in 10\,s, $1.0''$ rms in 60\,min \\
Tracking accuracy ($75^{\circ} < \mathrm{Alt} < 85^{\circ}$) & $0.4''$ rms in 10\,s, $3.0''$ rms in 60\,min \\
&    (degraded near zenith due to rapid azimuthal motion)\\
Maximum rotation speed & Az: $15^{\circ}\,\mathrm{s}^{-1}$, Alt: $5^{\circ}\,\mathrm{s}^{-1}$ \\
Dome  & Roll-off roof \\
                 \hline
\end{tabular}
 \end{center}
\end{table*}

\subsection{LATIOS: Large-FOV Transient Imager with CMOS}
\label{subsec:latios}

\begin{figure}
  \centering
  \includegraphics[width=8cm, angle=0]{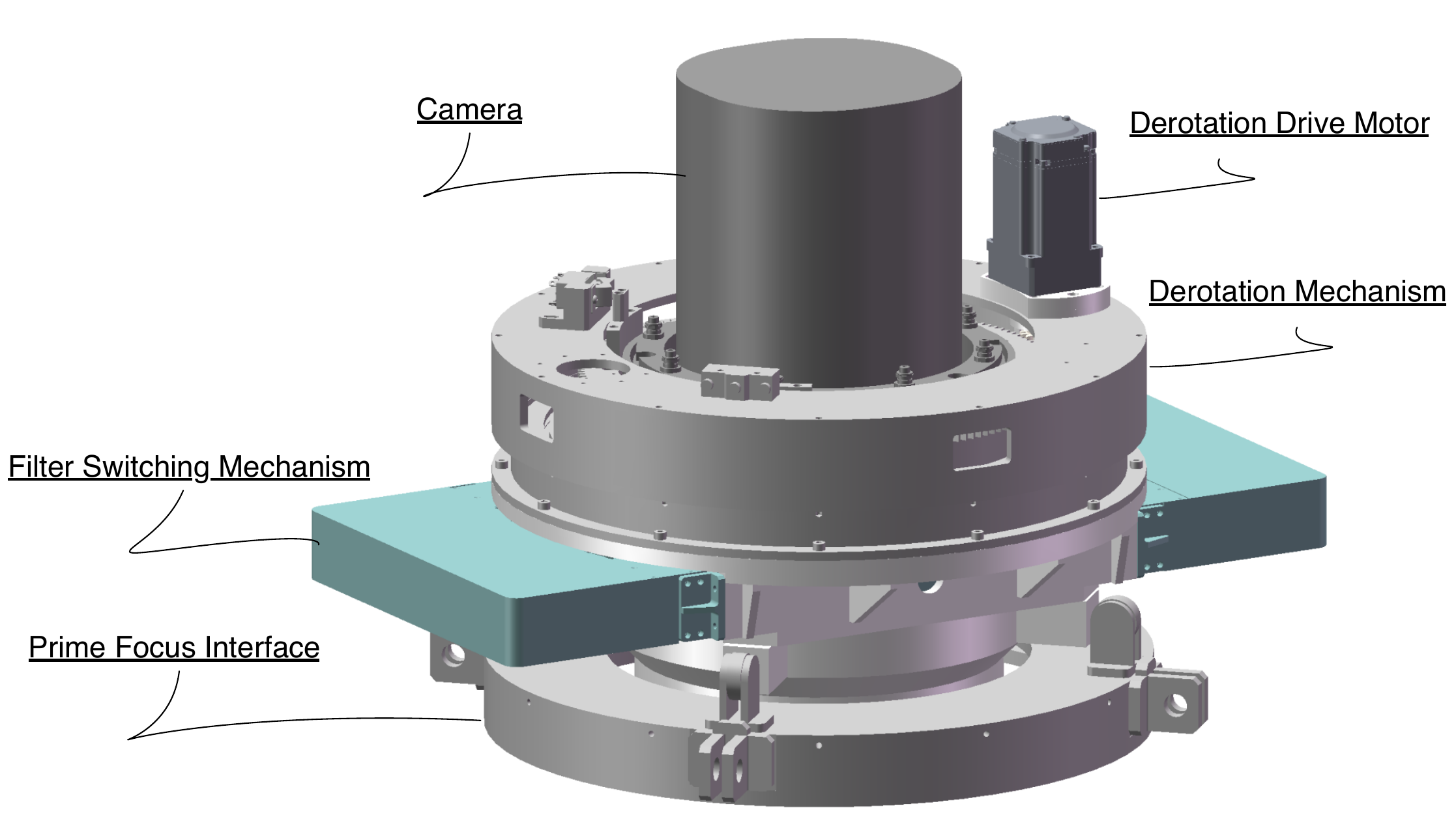}
  \caption{Schematic of the LATIOS assembly.}
  \label{latios_structure}
  \end{figure}

The prime-focus instrument, LATIOS, consists of a camera, a filter-switching module, and a field derotator (Figure \ref{latios_structure}). The camera employs a $4\mathrm{k}\times4\mathrm{k}$ sCMOS imaging sensor (Andor Balor 17F-12, $12\ \mu$m), providing an effective FOV\footnote{The quoted FOV corresponds to the nominal LATIOS configuration using an Andor Balor~17F-12 camera with a 12~µm pixel size. A backup, in-house–developed CMOS camera with 15~µm pixels is also available, providing an extended FOV of $\sim1.5^\circ \times 1.5^\circ$. However, it is still being extensively tested and is used only when specifically required.} of $1.28^{\circ}\times1.28^{\circ}$ with a pixel scale of $1.13''$/pixel. A notable advantage of the sensor is its large 70 mm diagonal, which affords wide-field coverage. It reaches a peak quantum efficiency (QE) of $\sim61\%$ near 600 nm and is free of etaloning across the entire response range, even close to the near-infrared, thereby ensuring accurate photometry. Its 18.5 ms full-frame readout\footnote{This value refers to the intrinsic sensor readout time; the effective frame interval in practice is typically longer due to data transfer, storage, and processing overhead.} supports time-domain astronomy of rapid phenomena, while low read noise ($<3\ e^-$) and excellent linearity ($>99.7\%$) enable precise light-curve measurements across a wide dynamic range. 

\begin{figure}
  \centering
  \includegraphics[width=8cm, angle=0]{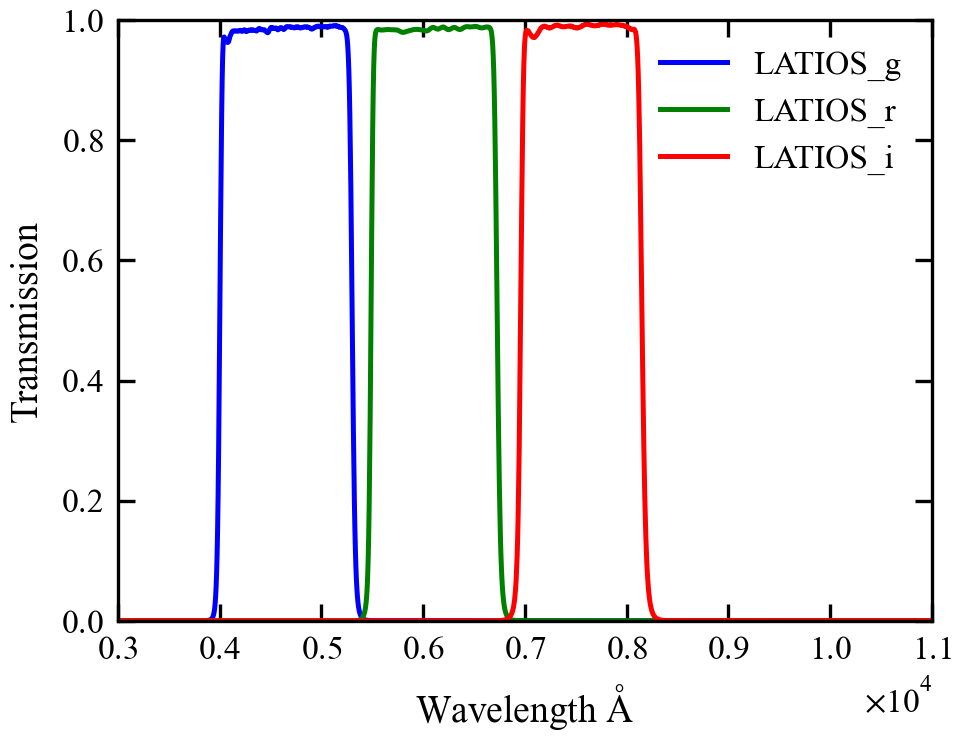}
  \caption{Transmission curves of the LATIOS all-dielectric filters (SDSS $g$, $r$, and $i$ bands) provided by the manufacturer Asahi Spectra Co., Ltd.}
  \label{latios_filters-gri}
  \end{figure}

\begin{figure}
  \centering
  \includegraphics[width=6cm, angle=0]{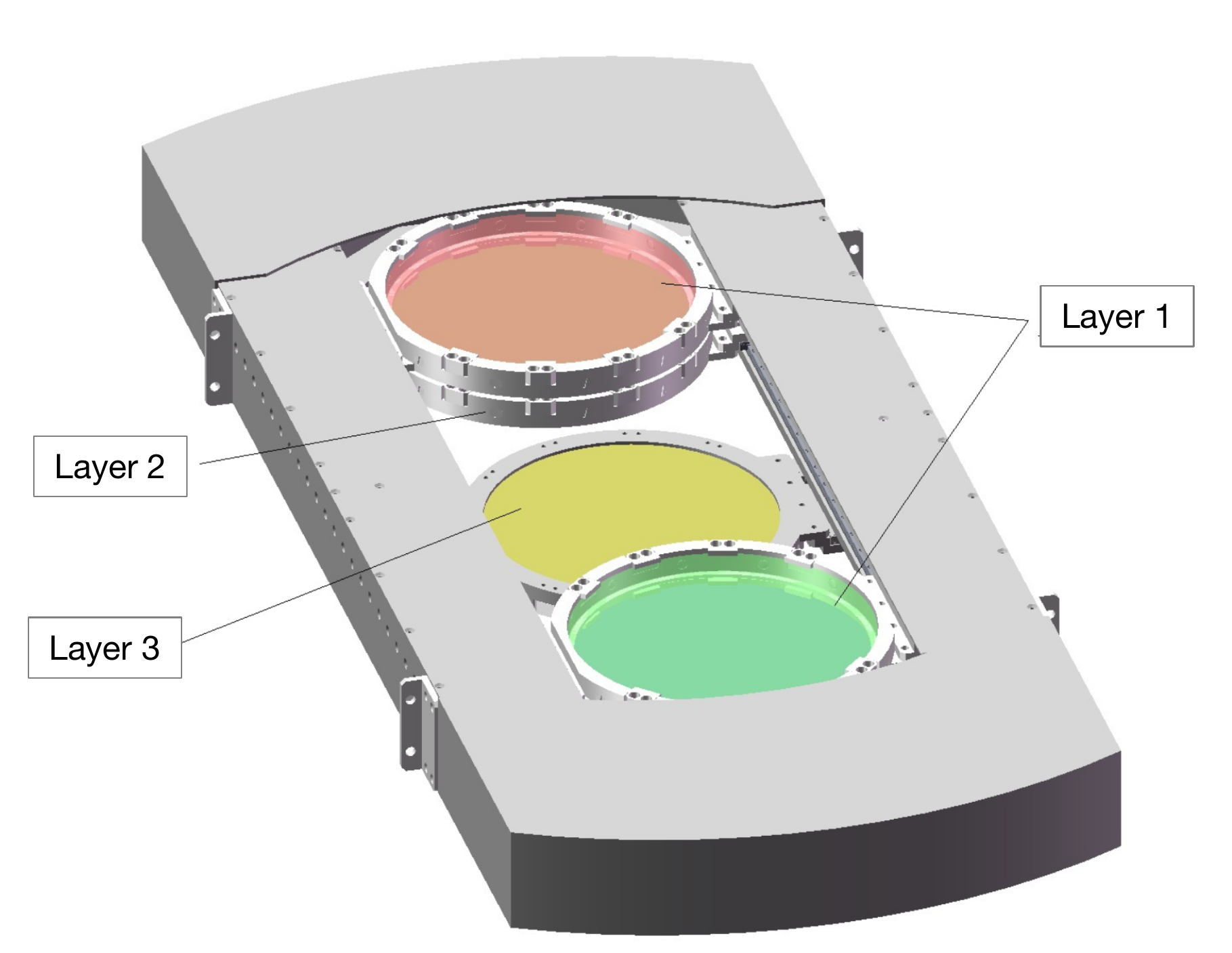}
  \caption{Layout of the LATIOS filter mechanism.}
  \label{latios_filters}
  \end{figure}

The filter-switching module facilitates multi-color observations in the SDSS photometric system. It is equipped with a set of high-transmission, all-dielectric SDSS filters\footnote{\url{https://www.asahi-spectra.com/opticalfilters/sdss-d.asp}} (Figure~\ref{latios_filters-gri}).
The three-layer, six-slot push–pull mechanism is designed for compactness, enabling motorized filter switching within a confined space and ensuring seamless transitions (Figure \ref{latios_filters}). 
The module accommodates the standard SDSS \(g\), \(r\), and \(i\) filters, a blank plate (for dark-field calibration), a fused silica flat (for white-light observations), and an empty slot reserved for future use. This configuration supports essential calibration procedures required by the prime-focus camera without a mechanical shutter.

The field derotator compensates for image rotation inherent to alt-azimuth mounts, ensuring stable tracking for unguided exposures exceeding five minutes. 
It utilizes Renishaw optical encoders that provide  $\sim1''$ feedback accuracy and employs dual read heads to suppress eccentricity errors.

The LATIOS control system consists of several independent control units for the camera, filter-switching mechanism, and field derotator, respectively, each capable of low-level standalone operation. 
These units are managed by a dedicated PC running a basic Linux operating system. 
An intermediate software layer, implemented in Python, provides an application programming interface (API) that enables users to issue fundamental control commands. 
Building upon these basic functions, users can develop extended command sets to perform complex operations such as telescope pointing and synchronized camera acquisition in response to alerts. 
The system can also automatically adjust the field derotator according to the telescope pointing to compensate for image rotation. The filter-switching mechanism achieves a transition time of less than 10~s, with an average of 7~s, enabling rapid multi-band observations.

Performance tests under moonless conditions indicate that LATIOS achieves a \(5\sigma\) detection limit of \(g=20.5\), \(r=20.6\), and \(i=20.0\)~mag in 180~s exposures,   
which satisfies the ground-based follow-up requirement of the \textit{SVOM} mission\footnote{Reaching \(\gtrsim 19.0\) mag within 5 min post-GRB trigger.}.
The system provides reliable multi-band imaging performance with minimal image rotation and accurate filter positioning, meeting the requirements for rapid-response and time-domain observations.

\subsection{CATCH: Camera with Three Channels}
\label{subsec:catch}
\begin{figure*}[!htbp]
  \centering
  \includegraphics[width=8cm, angle=0]{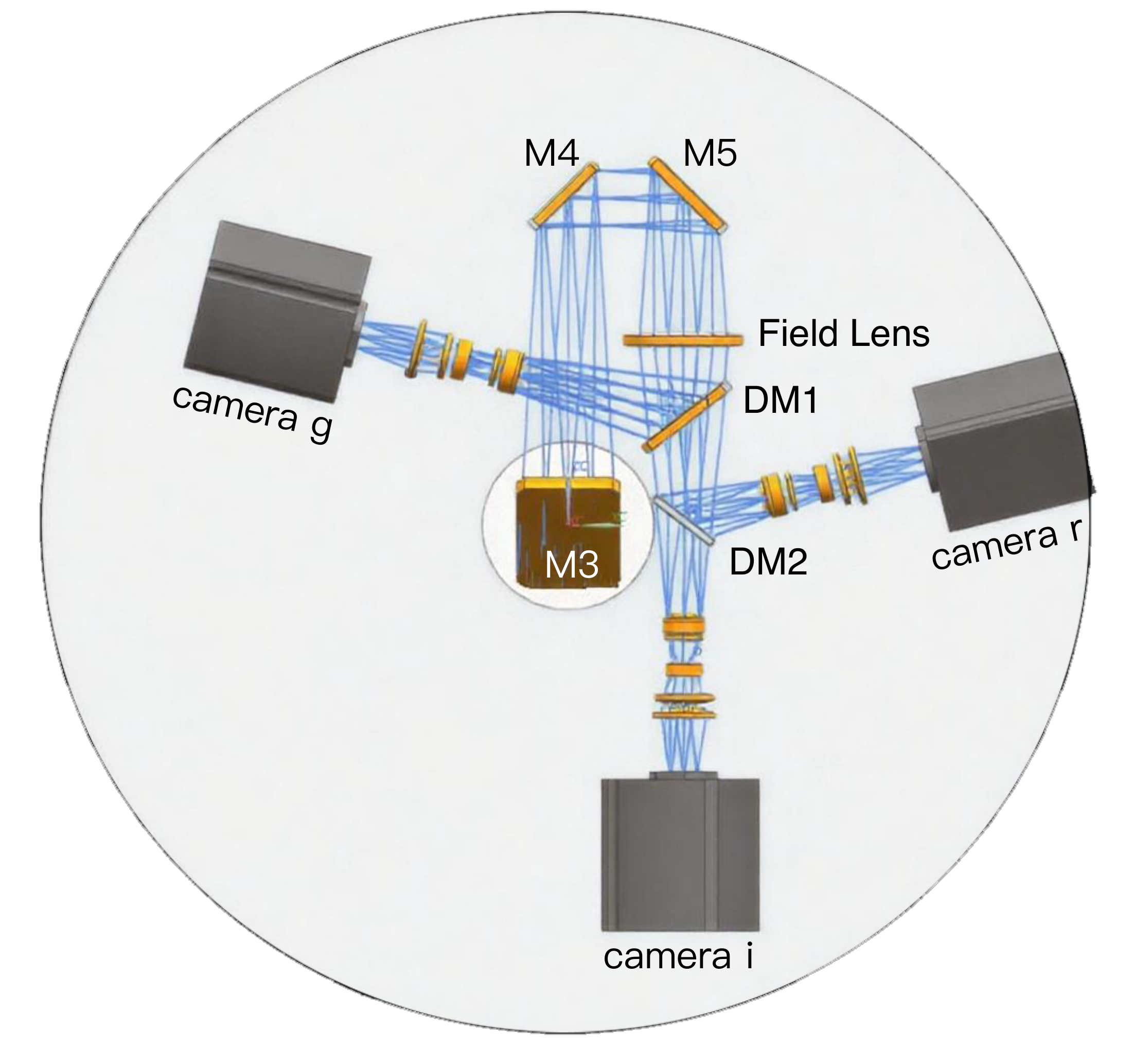}
  \caption{Optical layout of the CATCH system, showing redirection mirrors M3, M4, and M5, the common field lens, dichroic mirrors DM1 and DM2, and three $g$-, $r$-, and $i$-band cameras following their respective focusing lens groups.}
  \label{fig:optical_path_catch}
\end{figure*}

The optical configuration of the CATCH system is shown in Figure~\ref{fig:optical_path_catch}. Light collected at the telescope’s Cassegrain focus is redirected by the tertiary mirror (M3), folded by mirrors M4 and M5, and collimated by a field lens. The collimated beam is split by two dichroic mirrors (DM1 and DM2) into three optical channels corresponding to the \(g\), \(r\), and \(i\) bands. Each channel includes a dedicated lens group that reduces the focal ratio and corrects image aberrations, a respective SDSS bandpass filter (\(g\), \(r\), or \(i\)), and a detector, enabling simultaneous three-band imaging of a common FOV.

All optical components of the CATCH system are integrated within a terminal enclosure mounted behind the primary mirror cell. The enclosure is precisely positioned using locating pins to ensure high repeatability during assembly and disassembly, and it is secured with mechanical fasteners. Each optical group is equipped with adjustable mounts to allow fine alignment during integration. Given the relatively long effective focal length, several folding mirrors are employed to compact the optical path and to facilitate the alignment between each channel and the main optical axis. These mirrors are designed with three degrees of freedom, providing flexible and accurate adjustment capability during system calibration and maintenance.

\begin{figure}[!htbp]
  \centering
  \includegraphics[width=8cm, angle=0]{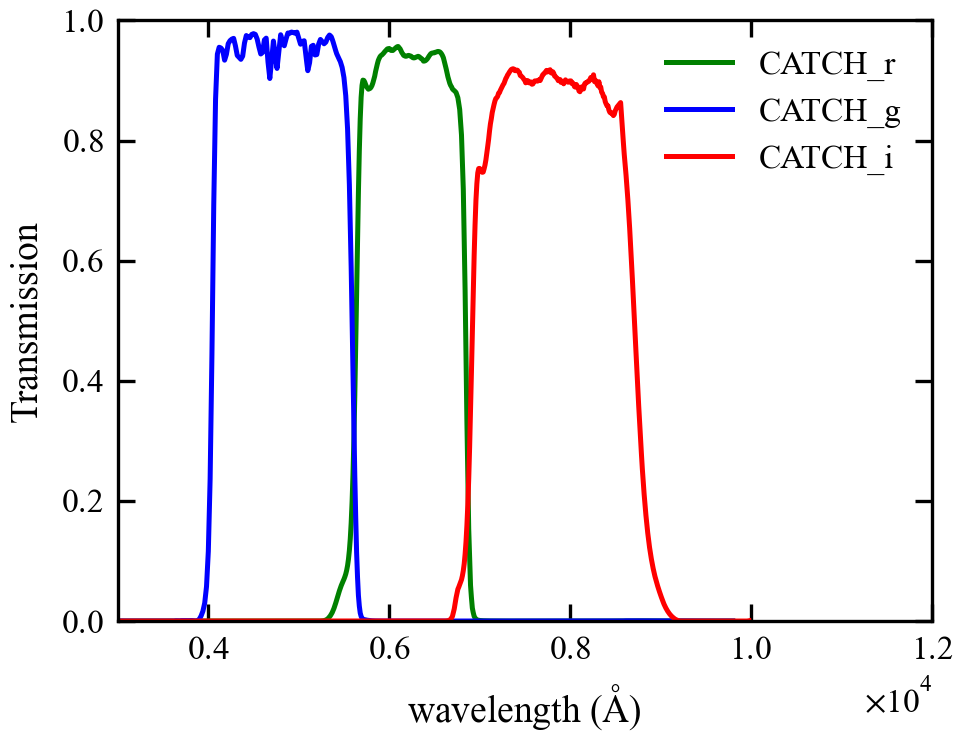}
  \caption{Transmission curves of the three CATCH channels, obtained by multiplying the filter and dichroic mirror transmissions.}
  \label{fig:channel_response_CATCH}
\end{figure}

All three channels of CATCH are equipped with Andor iKon-L~DZ936N-BV cameras. Each detector is a back-illuminated CCD detector with a \(2048 \times 2048\) array of 13.5~µm pixels, providing a \(21\arcmin \times 21\arcmin\) FOV. It employs thermoelectric cooling (TEC) down to --100~°C to suppress dark current and enable long-exposure, low-noise operation. Combined with a QE of up to 95\% at 500~nm, these characteristics make the camera well-suited for astronomical faint source observations. The transmission curves of the three CATCH channels are shown in Figure \ref{fig:channel_response_CATCH}.

\begin{figure}[!htbp]
  \centering
  \includegraphics[width=8cm, angle=0]{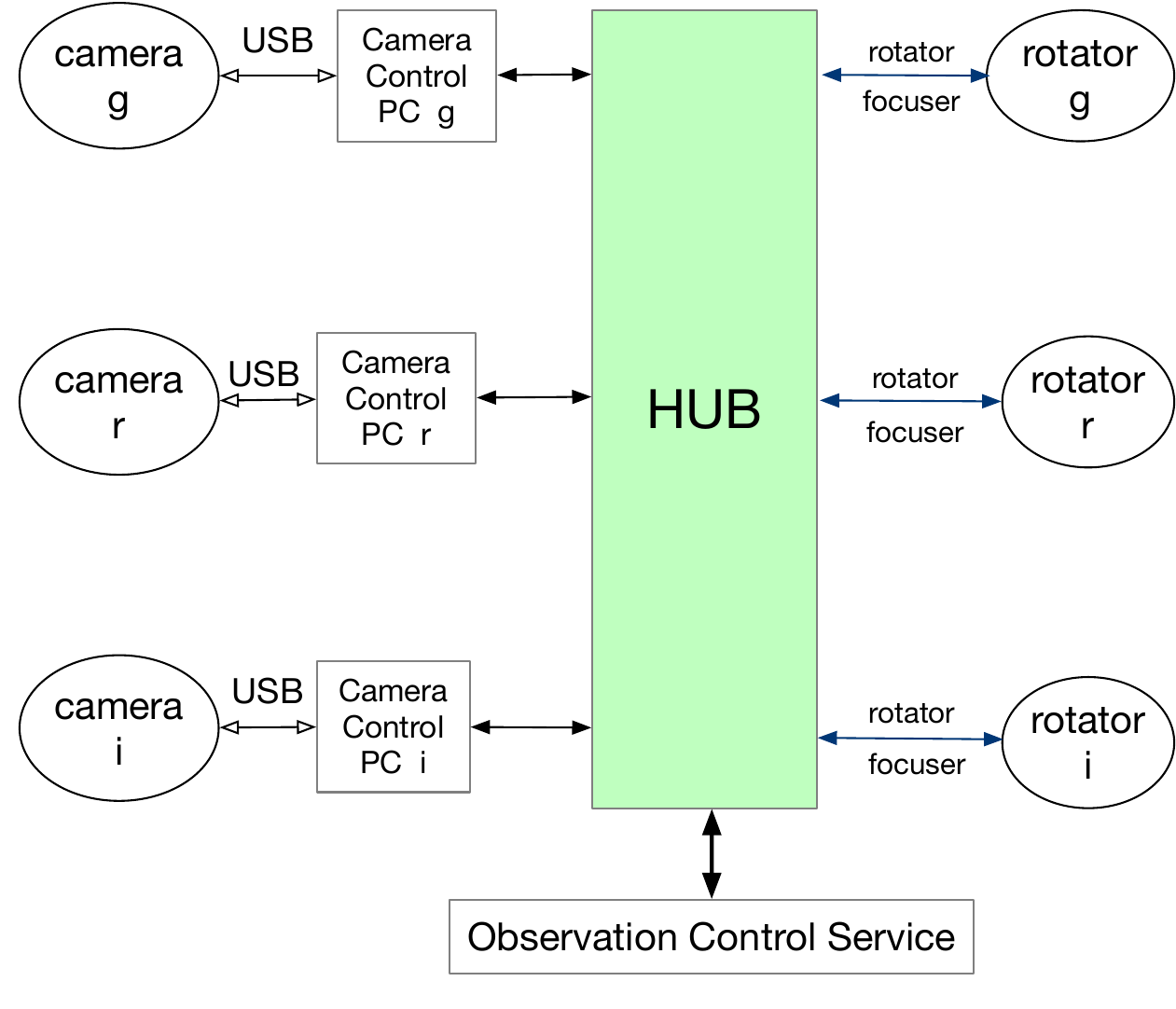}
  \caption{Schematic diagram of the CATCH control network architecture.}
  \label{fig:CATCH_control_cable}
\end{figure}

The control of CATCH is primarily managed by \texttt{camagent}, a C++--based software package, with the control network architecture displayed in Figure~\ref{fig:CATCH_control_cable}. 
The \texttt{camagent} operates the channel cameras and the de-rotator units\footnote{The Gemini Focusing Rotator manufactured by Optec provides both focusing and de-rotation capabilities.}, enabling image acquisition, real-time correction of field rotation by synchronizing the de-rotator with the telescope pointing, and focus position adjustment by the user. 
Each channel is controlled by an industrial computer running a Linux-based operating system; 
\texttt{camagent} is installed on this control computer, where it connects to the rotator via Ethernet and to the camera via USB. 
\texttt{camagent} provides a basic command set to external clients through a socket-based protocol. 
A Python API, referred to as the Observation Control Service and shared with the LATIOS main-focus system, interfaces with \texttt{camagent} and supplies a unified user-level communication layer for both instruments. 
Consequently, the operational command set for LATIOS and CATCH is largely identical, simplifying user-side control procedures and enhancing the system’s rapid response capability.

The observational performance demonstrates that CATCH meets its design specifications: under moonless conditions, it achieves a \(5\sigma\) detection limit of \(r \approx 19~\mathrm{mag}\) with an exposure time of 180\,s. 
The control system further enables a rapid response: once the telescope is slewed to the target position, it immediately initiates three-channel exposure acquisition without further delay.

\subsection{Operation}
\label{sec:operation}

\begin{figure*}
  \centering
  \includegraphics[width=12cm, angle=0]{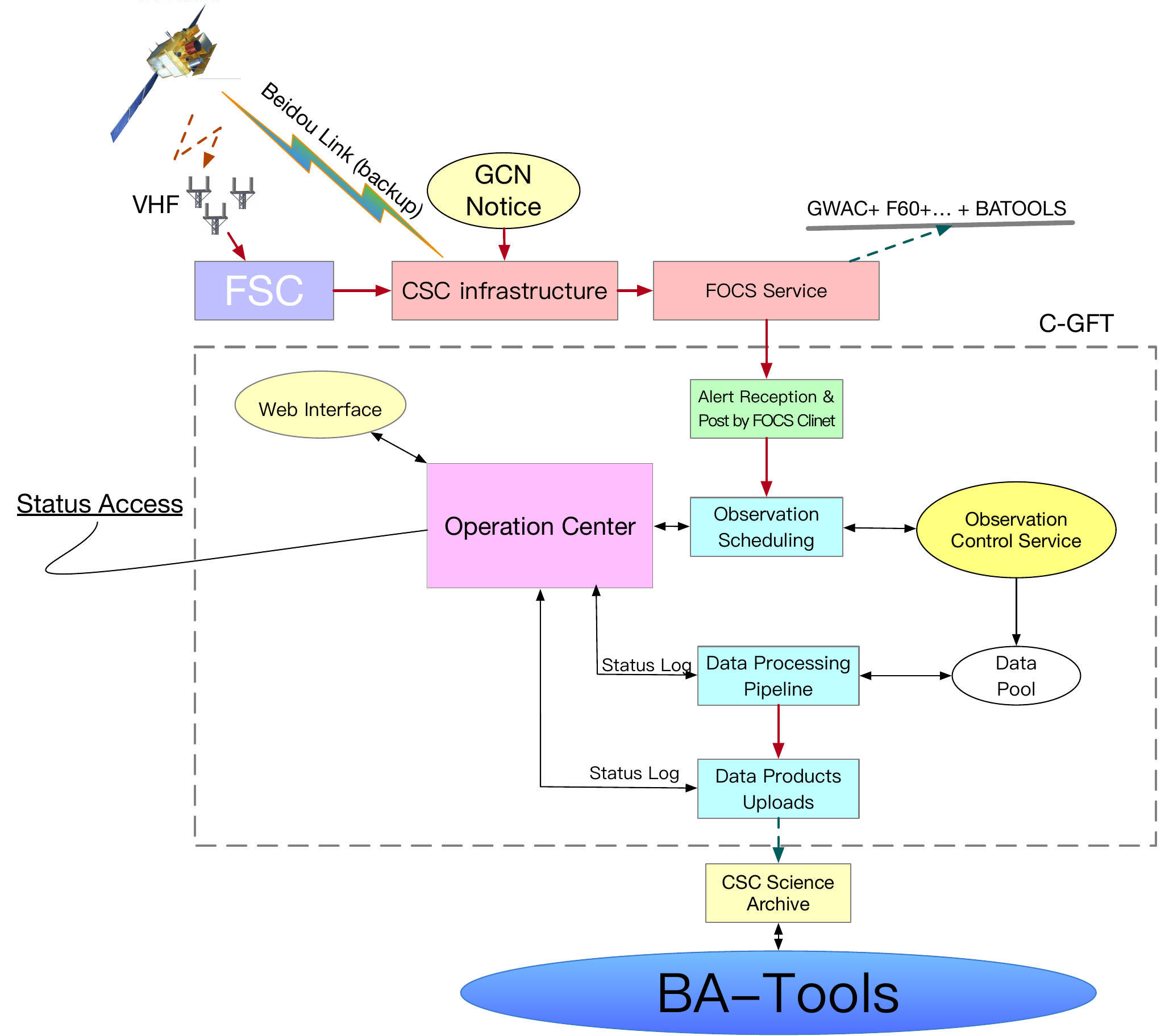}
  \caption{Workflow of the C-GFT automatic follow-up response.}
  \label{fig:workflow_cgft_followup}
  \end{figure*}

The overall fast-response workflow of the C-GFT is summarized in Figure~\ref{fig:workflow_cgft_followup}. The C-GFT receives transient alerts through the Follow-up Observation Coordinating Service (FOCS; \citealt{FOCShan2025svomfollowupobservationcoordinating}), operated at the \textit{SVOM} Chinese Science Center (CSC;\citealt{svomissue-BATOOLS-Xuhui2025svom}). Alerts generated by onboard \textit{SVOM} triggers are downlinked to the ground via the Very High Frequency ground antenna network (VHF; \citealt{svomissue-VHFlink-cordier2025svom})\footnote{The BeiDou short message service provides a backup channel for trigger downlink.} and then relayed to FOCS through the \textit{SVOM} network. In parallel, FOCS ingests external notices distributed by the The General Coordinates Network (GCN;\citealt{GCN2025HEAD...2240302R}),  such as alerts from the Einstein Probe (\textit{EP}; \citealt{EP2022hxga.book...86Y}) and the Neil Gehrels Swift Observatory (Swift; \citealt{swift2004ApJ...611.1005G}).

Upon receipt, each alert is decoded and standardized before being forwarded to the \textit{Observation Scheduling} module. Based on target visibility constraints, this module evaluates the feasibility of follow-up observations, computes the available observing window, and generates an observation plan according to predefined strategies. The plan specifies the execution timeline and command sequence, and supports both scheduled execution and interruption of ongoing observations, enabling fully automated and responsive operations.

At the scheduled time, the telescope executes the observation plan and stores the acquired data in a designated data pool. An online processing system continuously monitors this pool, triggers automatic data reduction upon the arrival of new files, and produces standardized \textit{SVOM} data products. These products are transferred to the CSC via HTTP Science Archive, where they can be visualized and interactively analyzed using the CSC BA Tools platform \citep{svomissue-BATOOLS-Xuhui2025svom}. In addition, a web-based interface allows users to track the execution status of observation plans and update schedules, while API services provide access to system logs and real-time monitoring of individual modules throughout the workflow.

Finally, the follow-up strategy leverages the complementary capabilities of the two focal-plane instruments. LATIOS offers a much larger field of view and higher sensitivity to faint sources, whereas CATCH provides simultaneous three-color imaging but with a shallower limiting magnitude. Since most newly detected optical counterparts are faint and high-energy triggers may have large localization uncertainties, LATIOS is used by default for the initial search and localization. If a bright counterpart has already been reported or the source is confirmed to be sufficiently bright, observations can be switched to the CATCH three-channel mode for simultaneous multi-band imaging. In practice, instrument switching is typically performed during daytime operations; therefore LATIOS is normally used for the first-night follow-up of newly triggered events.

\subsubsection{Alert Response and Operation Center}
\label{subsec:alert-response}

The dedicated FOCS  client deployed at C-GFT provides filtering of alerts and real-time monitoring of alert reception status, ensuring a rapid response to transient events such as GRB triggers from the \textit{SVOM} satellite or other external networks. Upon receiving external alerts from the FOCS Service, the Alert Receiver module of the FOCS client automatically validates and standardizes the alerts. The Alert Poster module of the client then forwards triggers to the \textit{Observation Scheduling} module after applying customizable filters (e.g., based on alert type or source instrument)\footnote{Currently, alerts from \textit{SVOM}, \textit{EP}, and \textit{Swift} can be selectively routed. Filtering rules are reconfigurable based on user requirements.}.

\begin{figure}
  \centering
  \includegraphics[width=8 cm, angle=0]{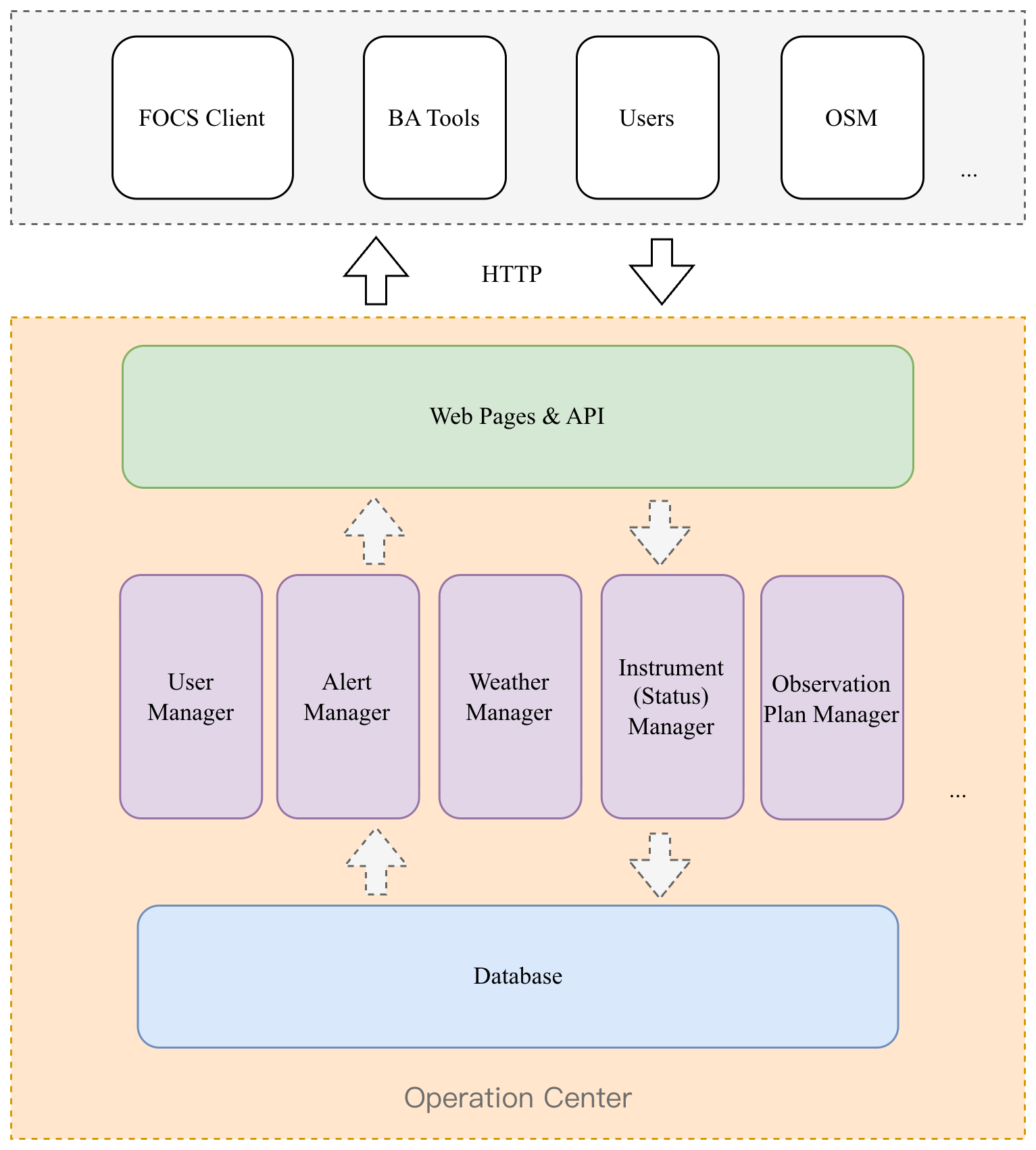}
  \caption{Overview of the C-GFT Operation Center architecture.}
  \label{figure:operation center}
  \end{figure}

As shown in Figure \ref{figure:operation center}, the Operation Center serves as the central hub coordinating the subsequent observation workflow. 
It interfaces with the \textit{Observation Scheduling Module} (OSM), part of the telescope and observation control subsystem (see Sect.~\ref{subsec:obs-control}), to dispatch observation tasks, while managing data processing, product uploads, and system status monitoring.
Communication with all subsystems, including the provision of status queries to users, is handled via the HTTP protocol. Through its web-based interface, the Operation Center also provides real-time access for both operators and external users, integrating automated alert response with routine observatory management.

\subsubsection{Telescope and Observation Control}
\label{subsec:obs-control}
\begin{figure*}
  \centering
  \includegraphics[width=14cm, angle=0]{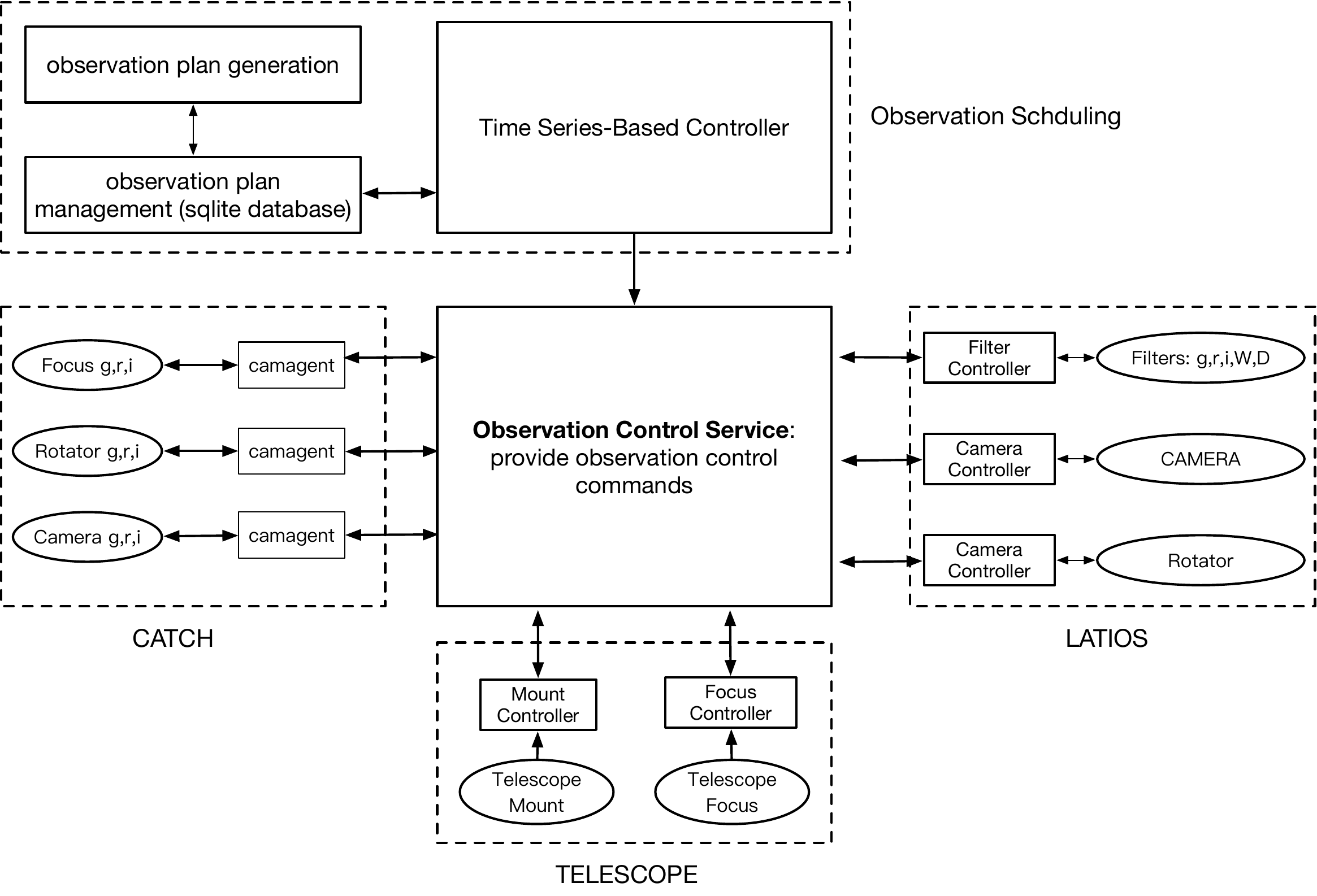}
  \caption{System architecture for the automated telescope and observation control.}
  \label{figure:OCS_TCS_flowchart}
  \end{figure*}

The Telescope and Observation Control functionality is primarily implemented through two modules shown in Fig.~\ref{fig:workflow_cgft_followup}: the \textit{Observation Scheduling} module and the \textit{Observation Control Service}. The detailed functions of these two modules are illustrated in the architecture of the automated telescope and observation control (Fig.~\ref{figure:OCS_TCS_flowchart}).
This architectural design was adopted because the C-GFT system was developed by upgrading an existing telescope with the LATIOS and CATCH hardware to enable rapid and fully automated responses to GRB alerts. Consequently, the control software was designed to maximize compatibility with the pre-existing operational framework.

The \textit{Observation Control Service} interfaces with three major hardware components: the telescope control unit and two observation terminals, and is primarily responsible for coordinating the operations between the telescope and the image acquisition terminals. The telescope control unit consists of the mount and the focusing subsystem, both of which retain their original low-level control modules. These modules communicate externally through a middleware service that provides socket-based interfaces\footnote{TCP for mount commands and UDP for focus status feedback}. Here, we refer to the LATIOS and CATCH units as observation terminals, each representing an instrument side control endpoint whose functionalities are exposed to the higher-level observation control through the \textit{Observation Control Service}. As shown in Fig.~\ref{figure:OCS_TCS_flowchart}, the LATIOS terminal provides socket interfaces via its middleware components, namely the \texttt{Camera Controller} and the \texttt{Filter Controller}. 
The CATCH terminal interfaces with the channel-level management modules (\texttt{camagent}) via socket-based interfaces to synchronize the operation of three identical channels, including control of the camera, focusing mechanism, and rotator. The calculation of the real-time de-rotation tracking is performed internally by each \texttt{camagent} and is transparent to the external control layer, which only enables or disables the tracking.

The \textit{Observation Control Service} integrates the command interfaces from the telescope control unit and the two observation terminals into a unified set of high-level functional commands, including telescope pointing, focusing, exposure start and stop, exposure time configuration, and filter selection (not required when using CATCH). The selection between the LATIOS and CATCH terminals is determined by configuration parameters\footnote{The selection is constrained by hardware considerations; switching the secondary mirror configuration requires daytime intervention.}.

The \textit{Observation Scheduling} module supports both automatic and manual observation modes. In automatic mode, it invokes the high-level command set through a \textit{Time-Based Controller}, which manages the timed execution and termination of tasks according to their assigned priorities. Higher priority tasks are placed in a waiting queue until their scheduled start time\footnote{During the waiting period, lower priority tasks may be executed.}, while expired tasks are skipped. Observation plan management is handled by a lightweight SQLite database.
The \textit{Observation Generation} module produces exposure plans based on incoming GRB alerts and predefined observing strategies. It calculates the observable time window from the target position and a minimum altitude constraint\footnote{Typically $>20^{\circ}$.}, and stores the corresponding observation instructions in the database. The \textit{Time-Based Controller} continuously monitors the database and assigns GRB observations the highest priority. When a new GRB observation plan is injected, the controller determines whether to initiate the observation immediately, wait until the target becomes observable, or skip the task if it is no longer executable.
In manual mode, observation commands can be issued directly through the terminal interface.

\subsection{Data processing}
\label{sec:data-processing}
The C-GFT data processing serves two primary objectives: (1) to rapidly identify the optical counterpart of a GRB within a localization  error of up to several hundred arcseconds provided by the high-energy instruments; and (2) to perform time-series photometry of the identified counterpart for scientific analysis.  In terms of timeliness, the system comprises two main pipelines: the C-GFT Quicklook Product Pipeline (CQPP), which focuses on automated processing and counterpart identification, and the C-GFT Refined (Standard Scientific) Product Pipeline (CRPP), which emphasizes interactive analysis, faint-source identification, and refined light-curve extraction. Both pipelines share the same set of fundamental calibration procedures.

\begin{figure}[htbp]
  \centering
  \includegraphics[width=8cm, angle=0]{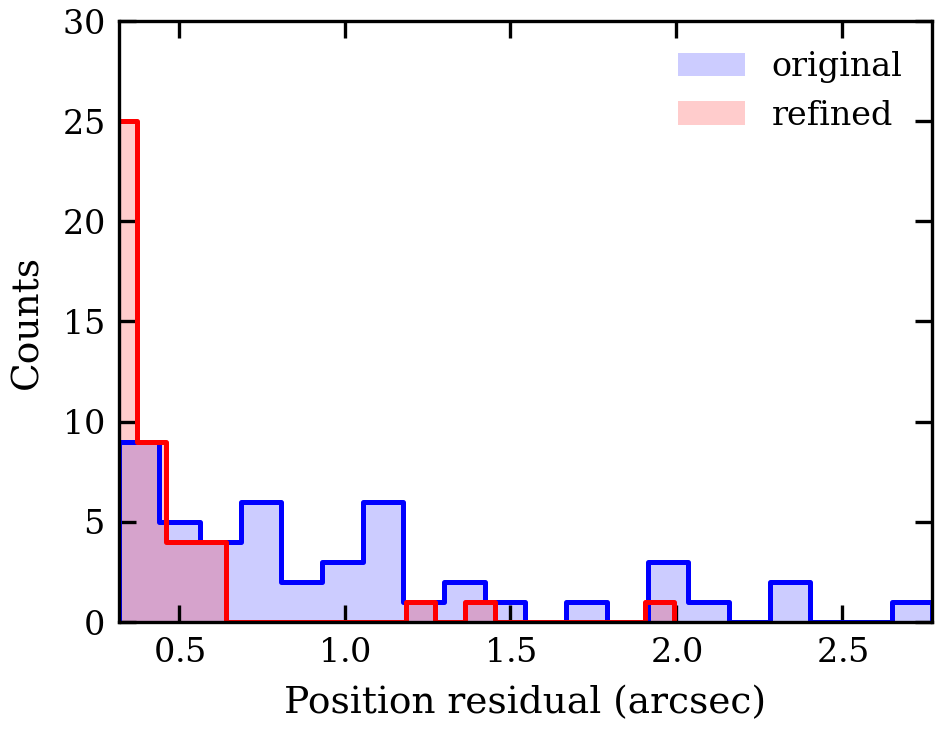}
  \caption{Distribution of astrometric position residuals (90\% C.L.) for 55 fields observed between August 2024 and November 2025, comparing the {\it original} \texttt{Astrometry.net} solutions with the {\it refined} results using the \textit{Gaia}~DR3 catalog.}
  \label{fig:hist_astrometry}
  \end{figure}

Astrometric calibration is performed using \texttt{Astrometry.net} \citep{astrometry.net2008ASPC..394...27H,astrometry.netlang2010} with a fourth-order polynomial solution. The reference index files are generated from the UCAC4 catalog \citep{ucac42013AJ....145...44Z}. To accelerate the processing, we constrain the pixel scale search range, allowing the initial astrometric solution to be completed within $\sim$8~s on average.
To improve astrometric calibration accuracy, refined processing is performed within a $26'\times26'$ region\footnote{The maximum localization uncertainty specified by the \textit{SVOM/ECLAIRs} design} using the \textit{Gaia} DR3 catalog \citep{gaiaDR32023A&A...674A...1G}. The positions of reference stars are further corrected for proper motion to the epoch of observation, with a fourth-order polynomial fit applied to matched sources to achieve precise celestial positioning.
We processed 55 fields observed on different nights from August 2024 to November 2025. 
Figure~\ref{fig:hist_astrometry} shows the distribution of the 90\% confidence level (C.L.) positional residuals. 
After refined calibration, the typical 90\% C.L. positional accuracy improves from 0.81\arcsec\ (original) to 0.37\arcsec. 
Inspection of the refined outliers indicates that all images with residuals greater than 1\arcsec\ correspond to 
severely defocused frames, and those with residuals above 0.5\arcsec\ exhibit noticeable image-quality degradation. 
Therefore, for images with normal quality, the refined astrometric accuracy is consistently better than 0.5\arcsec, 
with an average precision surpassing 0.37\arcsec\ (90\% C.L.).

Photometric calibration adopts the Pan-STARRS1 (PS1) catalog 
\citep{PS1survey-2016arXiv161205560C} as the flux reference standard. 
The PS1 catalog offers several advantages: 
(1) it provides nearly full coverage of all sky regions accessible to C--GFT; 
(2) it delivers high photometric accuracy, with a flux-calibration precision of 
7--12~millimagnitudes; and 
(3) its photometric system is closely tied to that of SDSS.
The LATIOS filter set follows the original SDSS design 
(Figure~\ref{latios_filters-gri}) and uses commercial high-transmission 
all-dielectric SDSS filters manufactured by Asahi Spectra Co., Ltd.
The CATCH instrument also adopts SDSS-like filters, although they were 
individually coated and therefore do not have identical transmission curves. 
In addition, CATCH uses a CCD detector, whereas LATIOS is equipped with a CMOS 
detector. Despite these hardware differences, analyses of the system response 
functions (Figure~\ref{fig:channel_response_CATCH}), together with on-sky tests 
reported in \cite{niu2022RAA....22e5009N}, demonstrate that the CATCH photometric 
system is also highly compatible with the SDSS system.
Given the close relation between the PS1 and SDSS photometric systems, 
color-term corrections are typically small \citep{PS1Calibration2012ApJ...750...99T}. 
Therefore, both LATIOS and CATCH adopt PS1 reference stars for absolute 
photometric calibration. For rapidly fading GRB optical afterglows, differential 
photometry can thus be performed directly using PS1 reference stars without 
applying a color term.

\begin{figure*}
  \centering
  \includegraphics[width=14cm, angle=0]{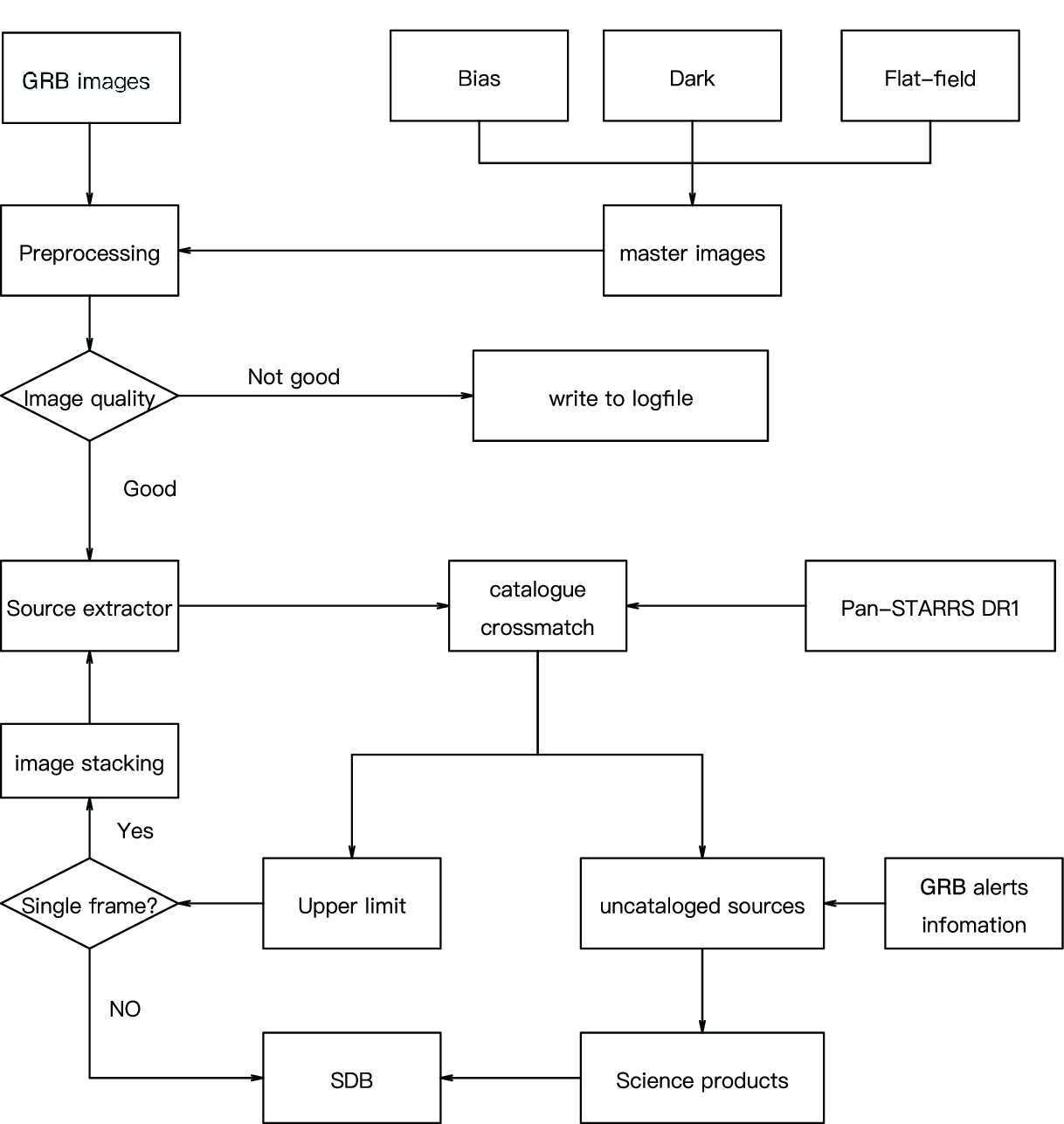}
  \caption{Flowchart of the C-GFT Quicklook Product Pipeline (CQPP). }
  \label{fig:data-pipepline-QuickProcessing}
  \end{figure*}

The CQPP is designed to rapidly reduce and automatically identify optical counterparts for GRBs from images obtained by C-GFT. As shown in Figure~\ref{fig:data-pipepline-QuickProcessing}, the processing begins with basic calibration, where raw GRB images are corrected using master bias, dark, and flat-field frames, which are regularly updated with calibration data. The preprocessed images are then subjected to a quality assessment step. Images that do not meet the quality criteria (e.g., poor tracking, excessive background, or low S/N) are automatically logged for further inspection, while those with acceptable quality proceed to the next stage.

For qualified images, sources are extracted using a standard detection algorithm, and the resulting source catalog is cross-matched with the Pan-STARRS DR1 reference catalog to identify known objects. By comparing the magnitude relationships between cataloged stars and observed stars, the flux calibration zero point is determined. This zero point can then be used to perform flux calibration for all stars in the observed catalog. Detections without catalog counterparts are flagged as uncataloged sources and temporarily stored for subsequent verification. The pipeline also estimates the upper-limit magnitude based on background noise statistics. Uncataloged sources are then cross checked with GRB alert information to identify afterglow candidates, which can be promptly reported for further confirmation.
If the processed image is a single frame image, it is aligned and combined with other observational images through image stacking to enhance the SNR and to detect faint transient. The source extraction and calibration process is then repeated for the stacked image, including flux calibration and verification steps.

The quick-look products include the original and calibrated images, the measured position and brightness of the GRB optical counterpart (if detected), and the upper-limit magnitude when no optical counterpart is found. 
All these products are packaged according to the \textit{SVOM} data product definition and uploaded to the scientific database of the Science Center for archival storage and further analysis.

\begin{figure*}[t]
  \centering
  \includegraphics[width=14cm, angle=0]{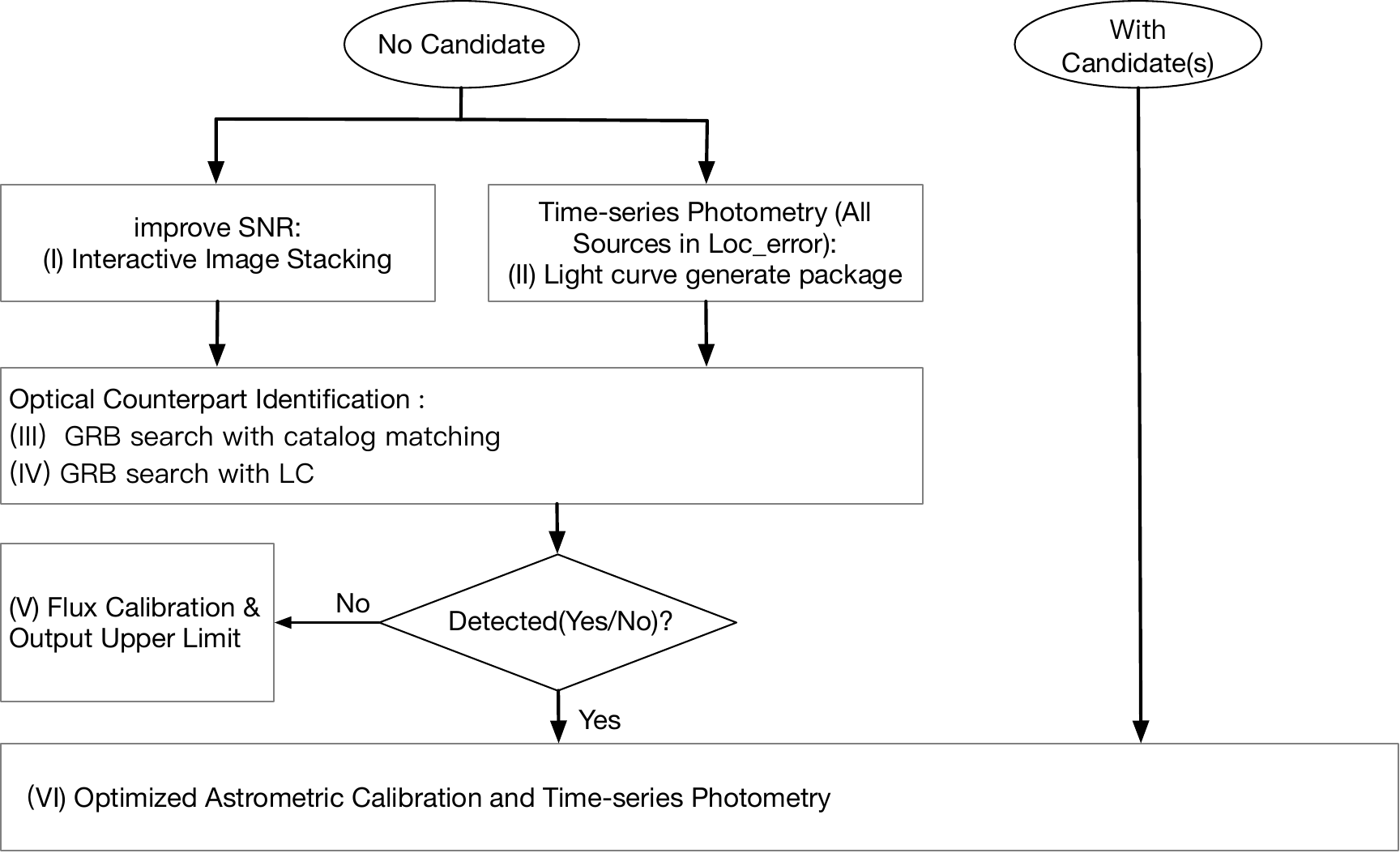}
  \caption{Flowchart of the C-GFT Refined (Standard Scientific) Product Pipeline (CRPP). }
  \label{fig:flowchart-CRPP}
  \end{figure*}

The CRPP provides an interactive environment for refined data processing, with its overall workflow illustrated in Figure~\ref{fig:flowchart-CRPP}. This workflow is structured around the two possible outcomes from the CQPP. If no optical counterpart candidate is found, the system performs: (1) interactive image stacking to improve the SNR for a deeper, catalog-matched search, and (2) a variability analysis of all stars within the trigger localization uncertainty to identify potential GRB counterparts based on their light curve morphology. 
If these refined searches yield no detection, the pipeline conducts photometric calibration and upper-limit estimation to generate standard \textit{SVOM} products.
Conversely, if a candidate is detected or a ``With Candidate(s)'' input is received, the pipeline executes optimized astrometric calibration and time series photometry. Finally, the refined coordinates and resulting photometric measurements of the GRB afterglow are packaged into standard \textit{SVOM} scientific products and delivered to the CSC.

Figure~\ref{fig:flowchart-CRPP} illustrates the six sequential processing modules (I--VI), each supported by a dedicated interactive data processing package.

I. Image Stacking.
This module interactively supports noise artifact rejection, background estimation, image quality assessment, and the selection of input frames for stacking. Its implementation is primarily based on the \texttt{geomap} and \texttt{geotran} tasks in IRAF \citep{IRAF1986SPIE..627..733T}.

II. Batch Light-Curve Generation.
Point-source detection and rapid aperture photometry are performed using SExtractor \citep{Sextractor1996A&AS..117..393B}. A relative photometric calibration is first established against a chosen reference image, followed by an absolute flux calibration of the reference frame itself. The module automatically produces light curves and manages all photometric time series data.

III. Source Extraction on Stacked Images.
After the SNR is improved through stacking, point sources are extracted and their photometry is performed. The resulting source catalog is cross-matched with the PS1 reference catalog, and any unmatched objects are flagged as potential optical counterparts for further inspection.

IV. Visual Identification of Light Curves.
This module enables visual examination of the light curves generated by Module~II to identify potential GRB optical counterparts. It also allows for the loading of candidate objects identified by Module~III, facilitating their further validation and analysis through corresponding light curves.

V. Photometric Upper-Limit Estimation.
For stacked images where no viable optical counterpart is detected, flux calibration is performed against the PS1 catalog. This enables the determination of a formal photometric detection limit for the observation.

VI. Optimized Astrometric and Photometric Calibration.
Astrometric refinement is performed using the Gaia DR3 catalog as the reference frame, a methodology detailed earlier in this section. The photometric optimization is specifically designed for faint sources and incorporates two complementary approaches:  
(1) Optimal Photometry: We implement the algorithm of \citet{optimalphoto1998MNRAS.296..339N} and \citet{optimalPhoto2002MNRAS.335..291N}, which extracts source flux using pixel wise weights derived from the point spread function (PSF) model and inversely scaled by the local noise variance. This technique is formally equivalent to profile fitting but achieves a higher SNR and exhibits reduced sensitivity to PSF modeling errors.  
(2) Fixed Position Photometry: After precisely aligning all time-series images to a common coordinate system, source positions are determined from the highest SNR frame. These coordinates are then held fixed to perform aperture photometry across all lower SNR images. This method is particularly well-suited for measuring faint, late time GRB afterglows.

\section{SYSTEM PERFORMANCE ANALYSIS}
\label{sec:performance-analysis}

\begin{figure}[htbp]
  \centering
  \includegraphics[width=8cm, angle=0]{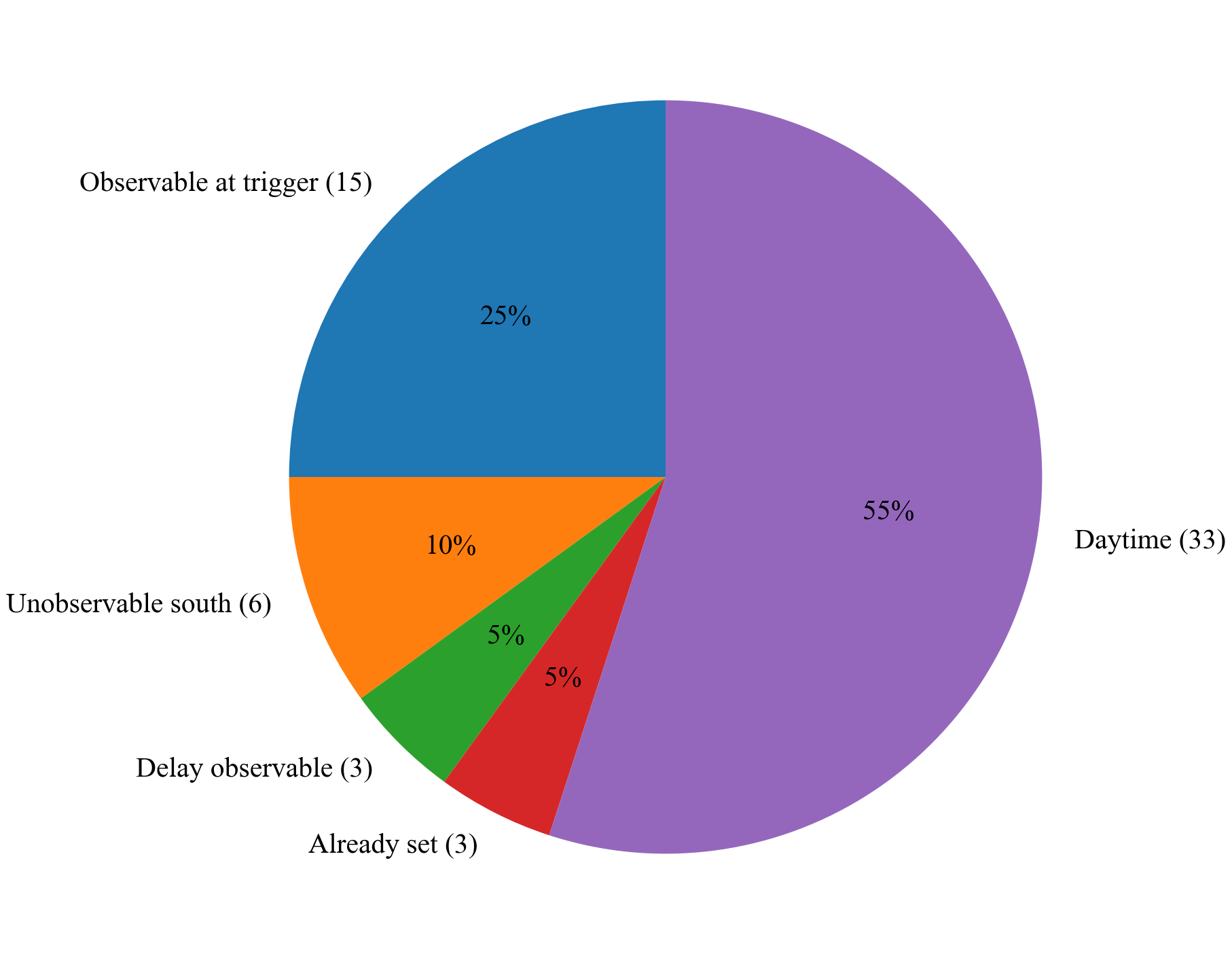}
  \caption{Distribution of \textit{SVOM/ECLAIRs} triggers by follow-up observability at the C-GFT site. Categories include daytime events, visibility-constrained nighttime targets (due to southern declination, already set, or delay), and events immediately observable at trigger time. The number of GRBs in each category is provided in parentheses.}
  \label{statistic_cgft_followup}
  \end{figure}

Figure \ref{statistic_cgft_followup} presents the performance statistics of C-GFT follow-up observations in response to \textit{SVOM/ECLAIRs} triggers over one year of commissioning and subsequent routine operations.
From the launch of \textit{SVOM} up to event sb25091206 \citep{sb25091206GCN41820} on 2025 September 12, a total of 60 GRBs were detected and localized by ECLAIRs, providing triggers suitable for follow-up by one or both GFTs. Among these, 27 events (45\%) occurred during nighttime at the C-GFT site. Of these 27, 15 events (25\% of the total) were located in the observable sky (defined as altitude $>20^{\circ}$ at trigger time). Regarding the remaining 12 nighttime triggers: 6 were unobservable due to their far-southern declination, 3 became accessible only after a delay, and 3 had already set at the trigger time (necessitating follow-up on the subsequent night). In total, C-GFT conducted rapid follow-up observations for 15 ECLAIRs-triggered events, corresponding to 25\% of all onboard ECLAIRs localizations.

We focus on the 15 trigger-time observable cases. The alert delivery chain proved stable with no failures in these cases: the latency from FSC alert generation\footnote{The generation time is extracted from the \texttt{voevent}.} to reception at the C-GFT ranged from 1 to 5~s (mean 2.7~s), and the subsequent command transmission from the alert reception system to the telescope control unit was likewise stable and uninterrupted. Of the 15 events, 10 were successfully observed by the C-GFT, while 4 were lost to weather and 1 to technical issues. Among the successful observations, adverse weather delayed the start of observations by 10--40 minutes in 3 cases, and in one instance, the ground based ECLAIRs trigger was received with a delay of $\sim$15~hr. For the 6 prompt responses, the latency between alert reception at C-GFT and the start of data acquisition ranged from 25 to 50~s, with a mean of 36~s, meeting the design requirement of less than one minute. Measured from the GRB trigger time, the observations started between 53 and 102~s,\footnote{One exceptional case, with a delayed FSC alert issued $\sim$9.7~min after the trigger time, is excluded from this range.} demonstrating C-GFT's capability for rapid follow-up. 

In addition to \textit{SVOM/ECLAIRs} triggers, C-GFT conducted prompt follow-up observations for 4 \textit{Swift} and 3 EP triggers, consisting of both GRBs and stellar flares. The automatic response system processes these external alerts in the same way as for \textit{ECLAIRs} triggers, and the telescope typically started observations within about one minute after receiving the alert, achieving a comparable response performance.

\begin{figure}
  \centering
  \includegraphics[width=8cm, angle=0]{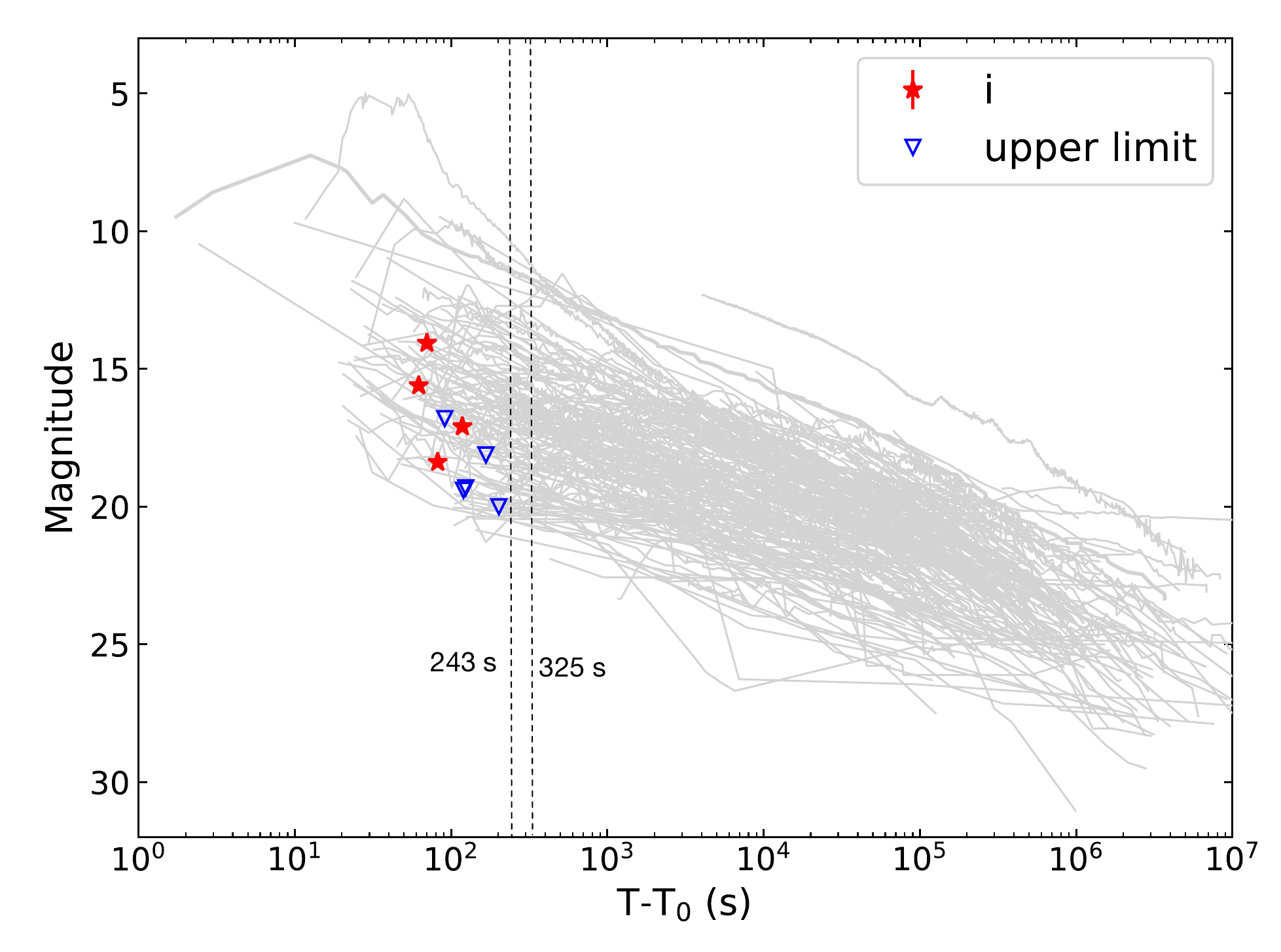}
  \caption{First C-GFT $i$-band detections (red stars) and upper limits (blue triangles) for 9 individual GRB triggers, plotted at the exposure mid-time relative to the trigger time $T_0$ (the 2nd and 3rd upper limits coincide). The data are overlaid on the historical optical afterglow light-curve compilation of \cite{2010ApJ...720.1513K,2011ApJ...734...96K} (gray lines). 
  The two vertical dashed lines denote the earliest (243 s) and median (325 s) start times of first-orbit VT observations after $T_0$, following automatic satellite slews.
  }
  \label{fig:cgft_followup_earlyphase}
  \end{figure}

As demonstrated in Figure \ref{fig:cgft_followup_earlyphase}, C-GFT successfully fulfills its assigned role within the \textit{SVOM} mission: observing early optical afterglows and bridging the temporal gap between the ECLAIRs trigger and the start of VT observations. The figure plots the first C-GFT $i$-band detections or 3$\sigma$ upper limits for 5 \textit{SVOM/ECLAIRs} GRBs and 4 \textit{Swift} GRBs that received prompt C-GFT follow-up. These data are overlaid on the historical optical afterglow light-curve compilation of \citet{2010ApJ...720.1513K,2011ApJ...734...96K}. 
The observations began between 62~s and 202~s post-trigger, where the times correspond to the mid-points of the exposures relative to the trigger time. Because the observing conditions varied among events, this temporal range also contributes to the detection capability under different environments. The achieved limiting magnitudes are typically $\sim 19$–20, obtained from stacked images of multiple 10~s exposures (typically $6\times10$~s and up to $18\times10$~s), balancing early-time coverage and depth.\footnote{Some early data points are from single 10~s exposures, in which the GRB optical counterparts were directly detected due to their high brightness.}
For comparison, the start times of the first-orbit VT observations following automatic satellite slewing during the past year of operations \citep{svomissue-VHFOnGround-Wu} are also indicated, with an earliest start time of 243~s and a median of 325~s post-trigger.

Benefiting from this fast response capability, C-GFT successfully captured early optical afterglow features, such as the light curve rise in GRB~250101A (76--150 s post-trigger; \citealt{GRB250101A2025GCN.38758....1S} ) and a prominent reverse-shock signal in GRB~240825A (65--180 s post-trigger; \citealt{GRB240825A2024GCN.37292....1S}). The latter suggests that the reverse shock of GRB~240825A is significantly more magnetized than the forward shock \citep{2025RAA....25j5003W}.

Regarding high-energy transients beyond GRBs, C-GFT detected the stellar flare EP~250523A. Multi-band observations of this event enabled an estimation of the white-light flare temperature through spectral energy distribution modeling (Wang et al. 2025, submitted).

\section{SUMMARY}
\label{sec:conclusion}
C-GFT is one of the two globally deployed ground-based follow-up telescopes\footnote{The other telescope, FM-GFT, was independently developed by the SVOM French team. While sharing the same scientific goals within the \textit{SVOM} mission, it adopts a different instrument design, featuring rapid response, multi-channel observations, and extended wavelength coverage up to the $H$ band ($\sim1.6\,\mu$m) in the near-infrared.} under the \textit{SVOM} mission, responsible for responding to alerts in the eastern hemisphere. It features a rapid response time (typically $<1$~minute), along with three-channel ($g$, $r$, $i$) and wide-field ($>1.28^\circ \times 1.28^\circ$) capabilities, though these two observing modes cannot be used simultaneously. This places it among the leading rapidly responding, multi-band facilities of comparable aperture worldwide, making it well suited for early-time GRB afterglow observations.
C-GFT conducts rapid multi-wavelength optical follow-up observations of \textit{SVOM} triggers, delivering early-phase, multi-wavelength afterglow observations. 
In particular, C-GFT provides critical early-phase GRB afterglow data that bridge the observational gap prior to the start of VT observations. The precise afterglow localizations obtained by C-GFT further enable timely spectroscopic and photometric follow-up with large ground based telescopes.
Beyond its core mission, C-GFT also responds to triggers from multi-messenger observatories including \textit{Swift}, the Einstein Probe, and the LIGO/Virgo/KAGRA network, enabling extensive time domain and multi-wavelength studies across a broad range of transients. More than one year of commissioning and subsequent routine operations demonstrate that C-GFT meets its design and performance requirements.

Future efforts will focus on improving the detection depth by mitigating background light contamination through optimized observing strategies and enhanced data processing techniques. In parallel, the observing modes and overall operational system will be further refined to strengthen robustness and long-term reliability.

\hrule

\begin{acknowledgements}
The Space-based multi-band astronomical Variable Objects Monitor (\textit{SVOM}) is a joint Chinese-French mission led by the Chinese National Space Administration (CNSA), the French Space Agency (CNES), and the Chinese Academy of Sciences (CAS). We gratefully acknowledge the unwavering support of NSSC, IAMCAS, XIOPM, NAOC, IHEP, CNES, CEA, and CNRS.

This work was supported by the National Key R\&D Program of China (Grant Nos. 2024YFA1611700, 2024YFA1611701 and 2024YFA1611702); the \textit{SVOM} project, a mission under the Strategic Priority Program on Space Science of the Chinese Academy of Sciences (CAS); and the CAS–Local Government Cooperation Project (No. 23SH04).
\end{acknowledgements}

\label{lastpage}
\clearpage
\bibliography{ms2026-0029}{}
\bibliographystyle{raa}  
\end{document}